\documentclass[12pt,oneside]{article}
\usepackage{amsfonts,amssymb,amsmath,here,graphicx}
\newcommand{\p}{\partial}

\newcommand{\be}{\begin{equation}}
\newcommand{\ee}{\end{equation}}
\newcommand{\ba}{\begin{array}}
\newcommand{\ea}{\end{array}}
\newcommand{\bea}{\begin{eqnarray}}
\newcommand{\eea}{\end{eqnarray}}
\newcommand{\beas}{\begin{eqnarray*}}
\newcommand{\eeas}{\end{eqnarray*}}

\def\0{\phantom{-}0}
\font\BB=msbm10
\def\RR{\hbox{\BB R}}

\def\TT{\hbox{\BB T}}

\begin{document}
\begin{center}
{\bf\sc\Large
Fourier law in a
momentum-conserving chain}\\
\vspace{1cm}
{Cristian Giardin\`a $^{\dagger}$ and Jorge Kurchan $^{\ddagger}$}\\
\vspace{1cm}
{\small $^{\dagger}$ EURANDOM} \\
{\small P.O. Box 513 - 5600 MB Eindhoven, The Netherlands}\\
{\small {e-mail: {\em giardina@eurandom.tue.nl}}}\\
\vspace{.5cm}
{\small $^{\ddagger}$ PMMH UMR 7636 CNRS-ESPCI}\\
{\small 10, rue Vauquelin}\\
75231, Paris CEDEX 05, France.
\vskip 1truecm
\end{center}
\vskip 1truecm
\begin{abstract}
\noindent
We introduce a family of   models for heat conduction with and without
momentum conservation. They are analytically solvable in the high temperature limit
and can also be efficiently  simulated.
In all cases Fourier Law is verified in one dimension.
\end{abstract}
\newpage


\section{Introduction}

When there is a temperature difference between the boundaries of a material, heat is
transported from the hottest to the coldest side. The
phenomenological law governing this  process has been known for a long time: the {\em Fourier law}
$J=k\nabla T$ states the
proportionality of the heat flux $J$ (the amount of heat
transported through the unit surface in unit time) to thermal
gradient $\nabla T$ (the spatial derivative of the temperature
field). The proportionality constant $k$ is called {\em thermal
conductivity coefficient}. Almost two centuries after  Fourier's
law was discovered, its microscopic derivation is still an open
problem from a fundamental point of view. At stake is not only a question
of mathematical rigor: spatial constraints in certain experimental systems
can significantly alter the transport properties in ways that are not yet
fully understood. Systems with dimension $d \leq 2$ can have a thermal
conductivity that can even become anomalous (i.e. diverging in the thermodynamical
limit), implying a breakdown of the usual Fourier law.
In the last few years such systems became experimentally realizable,
and this has spurred  a renewed  interest in the theoretical basis among
the physics community (for recent reviews see \cite{BLR:00,LLP:03}).

An important family of models which has been extensively studied
consist of regular periodic $d$ dimensional lattices of $N=L^d$ point-like
atoms interacting with their neighbors through {\em non-linear}
forces, as in the Fermi-Pasta-Ulam  (FPU) chain. 
For the linear chain ($d=1$) of oscillators it has been
rigorously established in the sixties \cite{RLL:67} that the heat
conductivity diverges in the thermodynamic limit like $k\sim L$.
This result was later generalized to the higher dimensional case
in \cite{N:70}.

Indeed, the dynamics of such a system can be decomposed into the evolution of
non-interacting waves (normal modes) which do not exchange energy
amongst each other, thus giving rise to ballistic transport. It
is nowadays clear that in order to have a finite thermal
coefficient the model must posess an efficient scattering
mechanism between phonons. However, numerical studies indicate
that also in the presence of anharmonicity heat conductivity is still
anomalous, with a power law divergence $k \sim L^{\alpha}$, where
$\alpha \sim 0.37$ \cite{LLP:97} for a linear chain. In recent years,
an increasingly large amount of (very often conflicting)
numerical results have appeared in the literature.
Although some progress have been achieved, many puzzles remain.
At present, the picture we have for 1$d$ is as follows:

\begin{itemize}

\item {\em Anomalous conductivity} occurs in cases in which momentum is
conserved, i.e. at least one acoustic phonon branch is present in
the harmonic limit \cite{LLP:97, H:99, D:01, GNY:02, CP:03,
DN:03}. The reason of the anomalous behavior has been traced
back to the low-frequency Fourier modes, which are only
weakly damped by the interaction with other modes
so they behave as very efficient energy carriers for
the system.
The condition of momentum-conservation is however not sufficient:
the `coupled rotor model' seems to have normal conductivity
\cite{GLPV:00, GS:00}, the reason being attributed to
phase jumps between barriers of the periodic potential
which act as scatters even for the long wave-length modes.

Here a short discussion is in order: if a system is composed of free
particles, momentum conservation is associated with the symmetry with respect to
simultaneous translation of all particles.
In a system such as the FPU or the coupled rotator system, we have particles with coordinates
$x_i$ -- a deviation from the lattice position in the former, and an angle in the latter case.
By translation invariance we may mean the symmetry `in the coordinates' $ i \rightarrow i+1$,
or the symmetry `in the fields' $x_i \rightarrow x_i + \delta$. The issue is somewhat mixed up
by the fact that in the FPU system we tend to picture the $x_i$ along the line joining the baths,
while for the coupled rotors we think of the $x_i$ as being transversal. In fact,
as already pointed out in \cite{NR:02}, the difference between `longitudinal' and `transversal'
only arises if the thermal baths interact with all the particles
 which reach a certain position
(so that there is contact with the bath or not depending on the actual value of $x_i$),
while it is irrelevant if the bath is  connected to  some fixed  particles at the end,
 whatever their positions $x_i$ might be. One could conjecture that
that the transverse nature of the $x_i$ in the rotor case
was responsible for the finite conductivity,
although this would suggest that the FPU chain with baths
coupled to fixed particles at the ends should
have normal conductivity too, contrary to observation.

\item {\em Finite thermal conductivity} has been obtained  for
special models with interaction with a substrate which
violates momentum conservation (the so-called Ding-a-ling
\cite{CFVV:84} and Ding-dong \cite{PR:92} models or various
Klein-Gordon chains \cite{HLZ:98, TBSZ:99}).

\item A Hamiltonian system for which a macroscopic
transport law has been rigorously derived is a {\em gas of
noninteracting particles} moving among a fixed array of periodic
convex scatters (periodic Lorentz gas or Sinai billiard)
\cite{BS:81,LS:78,AACG:99}. 
More recently, a modified Lorentz gas with fixed freely-rotating
circular scatters has also been considered \cite{MLL:01}.
In a further development, Eckmann and Young \cite{EY:04} introduced a related model
(having also normal heat conduction) that can be solved analytically in a  
limit.  

\item The consequences  of the underlying {\em microscopic dynamics}
are very controversial: while it was believed for a long time that
dynamical chaos and full hyperbolicity is a necessary and
sufficient condition in order to have  finite conductivity,
recent examples show that: i) deterministic diffusion and normal
heat transport can take place in systems with linear instability
\cite{ARV:02, LWH:02,LCW:03}; ii) there are systems with positive
Lyapunov exponents (FPU model) where the heat conduction does not
obey Fourier law.

\item The role of {\em disorder} (in the form of random masses of
the atoms) has also been considered. For the linear oscillators it
has been rigorously shown that disorder is not enough to reproduce
correctly Fourier law (in fact one finds $k\sim L^{1/2}$
\cite{CL:71, RG:71, OL:74}). For non-linear oscillators the
anomalous behavior has been confirmed numerically \cite{LZH:01}.

\item {\em Stochastic models} have also been studied
\cite{GKMP:81,KMP:82} and a finite thermal
conductivity has been found. The large deviation properties
of this kind of models have also been investigated recently
\cite{BGL:05}. While these models can be useful to obtain
exact results under the assumption of suitable Markovian dynamics,
one would like to obtain an explanation of Fourier law starting from
a pure deterministic description.

\end{itemize}

\noindent In the presence of anomalous heat conductivity
there exist mainly two general theoretical predictions:

\begin{itemize}

\item A self consistent `mode-coupling' theory which predicts an
exponent $\alpha = 2/5$ \cite{LLP:98,LLP:03}

\item Renormalization group analysis of  a set of
hydrodynamic equations for a $1$d fluid which yields $\alpha =
1/3$ \cite{NR:02}.
It has been suggested that a crossover might exist between these two
behaviours, see \cite{cinesi}.

\end{itemize}

\noindent Both of them apply to momentum conserving systems. It is
worth mentioning that present numerical simulations are not fully
in agreement with either mode-coupling or liquid theory
predictions. Some authors suggested that the discrepancy might
arise because intermediate sizes and timescales currently
available in simulations are not able to probe the real behavior
in the thermodynamic limit. In this respect, the problem must be
admittedly seen as an open question.The situation is even less
clear in $2d$ lattices. Theoretical arguments
support a weak logarithmic divergence for the thermal
conductivity. On the other hand numerical simulations
gave very conflicting results: logarithmic divergence
\cite{LL:00}, power law anomalous behavior \cite{GY:02}
and even finite thermal conductivity \cite{Y:02}.

To understand some of the basic features of the energy transport
in 1-$d$ systems, we develop in this paper a simple  model,
which can be cast also in the form of a set of discrete coupled maps such that
in a limit the continuum version is reobtained.
It can be solved analytically for high energies,
and it is suitable at all energies for numerical simulations because it is much less
computer-time demanding compared to a continuous  flux.
Because the approach to the high-energy value of the conductivity can be
seen in the simulations, the discussion is not plagued with doubts as to whether
the results are preasymptotic.
We have found no differences between the case in which momentum
is conserved and the case in which it is not.

\section{Hamiltonian model}

We consider a chain of particles for simplicity in a one-dimensional lattice.
In its most general form, the model is described by the following Hamiltonian
\be
H(x,p) = \sum_{i=1}^N \frac{1}{2}
\Big( p_i + A_i \Big)^2
\ee
where $A = (A_1(x),A_2(x),\ldots,A_N(x))$ is a
generalized `vector' potential in $\RR^N$
 and
$x = (x_1,x_2,\ldots,x_N)$,
$p = (p_1,p_2,\ldots,p_N)$
denote the particles position and momentum
respectively.
The Hamiltonian equations of motion read
\bea
\frac{dx_i}{dt}  & = & p_i + A_i  \nonumber \\
\frac{dp_i}{dt}  & = & - \sum_{j=1}^N \Big(p_j + A_j\Big)
\frac{\partial A_j}{\partial x_i}
\eea
They are more transparent in  Newtonian formulation
\bea
\label{newton}
\frac{dx_i}{dt} & = & v_i \nonumber \\
\frac{dv_i}{dt} & = & \sum_{j=1}^N B_{ij} v_j  +\frac{\partial A_i}{\partial t}
\eea
where
\be
B_{ij}(x) = \frac{\partial A_i(x)}{\partial x_j} -
\frac{\partial A_j(x)}{\partial x_i}
\ee
is an {\em antisymmetric} matrix containing in the
entries the `magnetic fields' acting on the particles.

\subsection{Conservation Laws}

{\em Conservation of Energy:}
Even if the forces depend on velocities and positions,
the model always conserves the total
(kinetic) energy if the fields are time-independent due to the
antisymmetry of the matrix $B$:
\be
\frac{d}{dt} \left ( \sum_i \frac{1}{2} v_i^2 \right ) =
\sum_{i,j=1}^N B_{ij} v_i v_j = 0
\ee
{\em Conservation of Momentum:}
Additional conserved quantities can be
imposed by a suitable
choice of the magnetic fields.
For example, if we choose the $A_i(x)$ such that they are left invariant
by  the simultaneous  translations $x_i \rightarrow x_i+\delta$, then the quantity
$\sum_i p_i$ is conserved. If in addition we require that $\sum_i A_i(x)=0 \;\; \forall x$,
then this also implies the conservation of $\sum_i \dot{x_i}$.

\section{Time-independent Hamiltonian, continuous time}

The simplest way to realize this is the following:
for a system with only energy conservation,
we may put $A_i=f_i(x_i,x_{i+1})$ for some functions $f_i$.
For periodic  boundary conditions, one identifies
the sites modulo $N$. Then, the matrix $B$ takes the
nearest-neighbour form:
\be
\label{b}
\mathbf{B} =
\left( \begin{array}{cccccc}
  \0   &   \phantom{-}B_1  &  \0      & \ddots   &  \0      &  -B_N   \\
 -B_{1\phantom{--}}  &   \0   &  \phantom{-}B_2     &   \0     & \ddots   &   \0    \\
  \0   &  -B_{2\phantom{--}}  &  \0      &   \phantom{-}B_3    &   \0     & \ddots  \\
\ddots & \ddots & \ddots   & \ddots   & \ddots   & \ddots  \\
\ddots & \ddots & -B_{N-3} &   \0     & \phantom{-}B_{N-2}  &   \0    \\
  \0   & \ddots &  \0      & -B_{N-2} &   \0     & \phantom{-}B_{N-1} \\
  \phantom{-}B_N  &   \0   & \ddots   &   \0     & -B_{N-1} &   \0    \\
\end{array} \right)
\ee
An open chain is obtained putting $B_N=0$.
A momentum-conserving model with next-to-nearest neighbor interaction can be
obtained with:
\begin{eqnarray}
A_i&=&C_{i+1}-C_{i} \nonumber \\
C_i&=& f_i(x_{i+1}-x_{i})
\end{eqnarray}
for any functions $f_i$. Again, one can take periodic boundary conditions
(and then the indices are interpreted as modulo $N$), or consider an open chain,
in which case one has to make $C_1=0$.

In both cases the end sites may be coupled to Langevin baths by modifying the equations of motion
of $p_1$ and $p_N$ adding a noise and a friction term, satisfying the fluctuation-dissipation
relations for temperatures $T_1$ and $T_N$, respectively.

\section{Discrete time}

One can also consider  discrete-time versions of these models, yielding a  system of coupled maps.
These have the advantage of being particularly easy to simulate, and, more importantly,
they are analytically solvable in the high energy limit.
As we shall see, the price we shall pay is that the map will not be simplectic \cite{sym}, but it does preserve
the phase-space volume  and the Gibbs measure.

We first describe a version without momentum conservation in detail, and then
describe more briefly the momentum-conserving case.
We consider impulsive magnetic fields
which act periodically and are of the form:
\be
B_{i,j} (x,t) = B_{ij}^{even}(x) K^{even}(t) + B_{i,j}^{odd}(x) K^{odd}(t)
\ee
where
\begin{eqnarray}
K^{even}(t) &=& \sum_{n=-\infty}^{+\infty} \Delta(t-n\tau) \nonumber \\
K^{odd}(t) &=& \sum_{n=-\infty}^{+\infty}\Delta(t-(n+1/2)\tau)
\end{eqnarray}
where $\Delta(u)$ is a square pulse of intensity $k$ and duration $1/k$, and we consider $k \rightarrow \infty$.
During the short time when the  fields are acting, the particles move a negligible amount,
but their velocities rotate under their action. On the contrary,
between kicks the velocities are constant and the motion is uniform.
Without loss of generality we can set $\tau = 1$ because
changing the time interval between kicks amounts to
rescaling the velocities.
In the spirit of Refs. \cite{GKMP:81,KMP:82,EY:04}, we consider a chain with even $N$ and
 choose the $B_{ij}^{even}$ and $B_{ij}^{odd}$ such that the impulsive fields
couple  sites pairwise as follows:
\begin{itemize}
\item $B_{ij}^{odd}$ is of the form (\ref{b}) with $B_2=B_4=B_6= ...=0$, and
   $B_1=G(x_1,x_2)$, $B_3=G(x_3,x_4)$, ..., etc.
\item
$B_{ij}^{even}$ is of the form (\ref{b}) with $B_1=B_3=B_5= ...=0$, and
   $B_2=G(x_2,x_3)$, $B_4=G(x_4,x_5)$, ... etc.
\end{itemize}

A periodic chain might be used, or, as we shall do later, one can
connect the end sites to thermal baths  during a half-cycle between kicks.
The definition is completed by saying that whenever a site is connected to a heat bath
during half a cycle, the interaction is so strong as to completely equilibrate
the site at the bath's temperature.

As we mentioned above, the map as we define it is not symplectic, 
even though it might seem as the composition of Hamiltonian steps.
The reason for  this is subtle: just before the end of the period during which the
 magnetic is on, the velocity is $v_i = p_i + A_i$, and just after it is turned off it is $v_i'=p_i$.  
If we wish to impose that $v_i=v_i'$ (which we need for energy conservation), we see that this implies a jump
in the momenta of magnitude $A_i(x)$. This is a transformation that is {\em not} symplectic unless
$\frac{\partial A_i}{\partial q_j}=\frac{\partial A_j}{\partial q_i}$, as is easy to check.
 An alternative way to see this issue is to consider the map as resulting from a time-dependent field, and
then adding a (nonconservative!) force that exactly cancels $\frac{\partial A_i}{\partial t}$ in (\ref{newton}),
so that energy is conserved. 
Note, however, that the map conserves the volume for all trajectories, and leaves the Gibbs measure invariant,
even if there is a potential $V(q)$  acting during the intervals between kicks.

Because in this time-dependent version we have given up symplecticity, we might just as well give up
the requirement that the $B_{ij}=B_{ji}$ derive from a `vector potential' $A_i$.
 This will neither alter the equilibrium nor
the volume conservation properties of the map, but gives us the freedom to use simpler functional forms for the
$B_{ij}$.   We tried different possibilities for the
$x$-dependence of the magnetic field acting on each couple
of sites, obtaining similar
results.
 We present here those corresponding to the choice:
\be
\label{magneticfield}
G(x_i,x_{i+1}) = K [ (x_i + x_{i+1}) - 2 \pi]
\ee
With this choice the magnetic fields take values
in the interval $[-2 K\pi, 2 K\pi]$, where $K$ is a
constant parameter.

\noindent
By integrating the equations of motion (\ref{newton})
between  two successive kicks, and having imposed the condition of continuity of the velocities described above,
 we arrive at the following
discrete time dynamics: denote by $R_i(t)$  the $2\times2$
rotation matrix
\be
R_i(t) = \left (
\ba{cc}
 c(B_i) & s(B_i) \\
-s(B_i) & c(B_i)
\ea
\right )
\label{cosa}
\ee
where
\bea
c(B_i) & = & \cos (B_i(t))
\nonumber \\
s(B_i) & = & \sin (B_i(t))
\eea
and consider the matrices $C(t)$ and $D(t)$
composed of a series of two by two blocks:
\be
C(t) =
\left(
\begin{array}{c|c|c|c|c|c}
R_1
&
\ba{cc} 0 & 0 \\ 0 & 0 \\ \ea
&
\ba{cc} 0 & 0 \\ 0 & 0 \\ \ea
&
\cdots &
\ba{cc} 0 & 0 \\ 0 & 0 \\ \ea
&
\ba{cc} 0 & 0 \\ 0 & 0 \\ \ea
\\
\hline
\ba{cc} 0 & 0 \\ 0 & 0 \\ \ea
&
R_3
& & & &
\ba{cc} 0 & 0 \\ 0 & 0 \\ \ea
\\
\hline
\ba{cc} 0 & 0 \\ 0 & 0 \\ \ea
& &
R_5
& & &
\ba{cc} 0 & 0 \\ 0 & 0 \\ \ea
\\
\hline
\vdots & & &
\ba{cc} \ddots & \\ & \\ \ea
& & \vdots \\
\hline
\ba{cc} 0 & 0 \\ 0 & 0 \\ \ea
& & & &
R_{N-3}
&
\ba{cc} 0 & 0 \\ 0 & 0 \\ \ea
\\
\hline
\ba{cc} 0 & 0 \\ 0 & 0 \\ \ea
&
\ba{cc} 0 & 0 \\ 0 & 0 \\ \ea
&
\ba{cc} 0 & 0 \\ 0 & 0 \\ \ea
&
\ldots
&
\ba{cc} 0 & 0 \\ 0 & 0 \\ \ea
&
R_{N-1}
\\
\end{array}
\right)
\ee
\be
D(t) =
\left(
\begin{array}{c|c|c|c|c|c|c}
0
&
\ba{cc} 0 & 0 \\ \ea
&
\ba{cc} 0 & 0 \\ \ea
&
\cdots
&
\ba{cc} 0 & 0 \\ \ea
&
\ba{cc} 0 & 0 \\ \ea
&
0
\\
\hline
\ba{c} 0 \\ 0 \ea
&
R_2
& & & & &
\ba{c} 0 \\ 0 \\ \ea
\\
\hline
\ba{c} 0 \\ 0 \ea
& &
R_4
& & & &
\ba{c} 0 \\ 0 \\ \ea
\\
\hline
\vdots & & &
\ba{cc} \ddots & \\ & \\ \ea
& & & \vdots \\
\hline
\ba{c} 0 \\ 0 \ea
& & & &
R_{N-4}
& &
\ba{c} 0 \\ 0 \\ \ea
\\
\hline
\ba{c} 0 \\ 0 \ea
& & & & &
R_{N-2}
&
\ba{c} 0 \\ 0 \\ \ea
\\
\hline
0
&
\ba{cc} 0 & 0 \\ \ea
&
\ba{cc} 0 & 0 \\ \ea
&
\cdots
&
\ba{cc} 0 & 0 \\ \ea
&
\ba{cc} 0 & 0 \\ \ea
&
0
\\
\end{array}
\right)
\ee

\vspace{0.5cm}
\noindent
Given $x(t),v(t)$, the position-velocity vector at time $t$,
its evolution will be
\bea
\label{one}
x(t+1/2) = x(t) + v(t) \nonumber \\
v(t+1/2) = C(t) \cdot v(t)
\eea
\bea
\label{two}
x(t+1) = x(t+1/2) + v(t+1/2)\nonumber \\
v(t+1) = D(t) \cdot v(t+1/2) + \xi(t)
\eea
where $\xi(t)$ is a $N$-dimensional
vector which is introduced to model
interaction with heat baths
\be
\xi(t) = \left (
\ba{c}
\xi_1(t) \\
0\\
0\\
\vdots \\
0 \\
0 \\
\xi_N(t) \\
\ea
\right )
\ee
At each time the non-zero components $\xi_1(t)$
and $\xi_N(t)$ are i.i.d. random variables
with Gaussian distribution
\be
P(y) = \frac{1}{\sqrt{2\pi \sigma^2}}
e^{-\frac{y^2}{2\sigma^2}}
\ee
where the variances of $\xi_1(t)$ and $\xi_N(t)$ are
respectively $T_1$ and $T_N$.

All odd sites interact with the site to their right during half a cycle, and with the site
to their left during the other half cycle.
The sites at the ends interact with the baths during the corresponding half-cycles, and
during that time they completely thermalize.
 In order to keep things simple we restrict the
configuration space of each particle to the torus
($x\in[0,2\pi]$), so that all the $x_i(t)$ are understood modulo $2\pi$.

As usual with these maps, if we consider the limit of small
velocities and weak kicks, we recover a continuous evolution.

\subsection{Chaoticity properties of the map.}

\begin{figure}
\begin{center}
\includegraphics[width=15.cm]{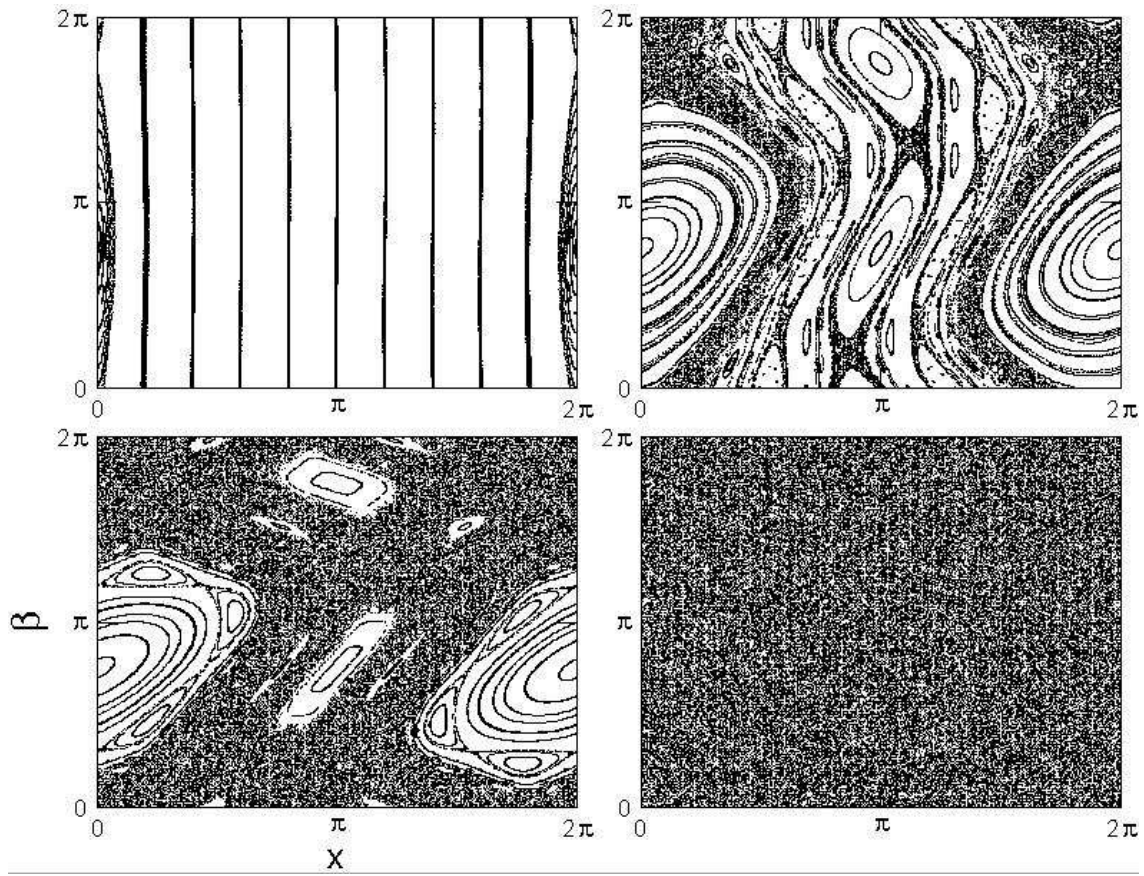}
\label{poincare}
\caption{{\small
Poincare sections for the map (\ref{map}).
Here $K=1$ and $v=0.01$ (top left), $v=0.6$ (top right),
$v=1.$ (bottom left), $v=5.$ (bottom right)}}
\end{center}
\label{fig1}
\end{figure}

In order to see what chaoticity properties to expect from a chain, we
begin in this section with the study of the dynamical properties of
the elementary map with only two sites $(x_1,x_2) \equiv (x,y)$.
This a particle moving in the $x-y$ plane 
(actually we restrict ourselves to the torus $\TT^2$)
under the action of an impulsive  field along the $z$ direction
whose amplitude depends on the particle's position itself.
Between kicks the motion is free; at each kick the velocity vector is
rotated by an angle which is  exactly given by the magnetic field,
evaluated at the point where the kick takes place.
Since the dynamics conserves the energy
the accessible phase space is 3-dimensional.
Denoting by $v$ the modulus of the velocity
$(v = \sqrt{v_x^2 + v_y^2})$
and by $\beta$ its angle w.r.t.
the $x$ direction in the plane,
we have
\bea
\label{map}
x (t+1)      & = & \{x(t) + v \cos (\beta(t))\} \mod(2\pi) \nonumber \\
y(t+1)      & = & \{ y(t) + v \sin (\beta(t)) \}\mod(2\pi)  \nonumber \\
\beta(t+1)  & = & \beta(t) + K[ ( x(t+1)+y(t+1) )  -2\pi]
\eea
The two-site map has two control parameters: the modulus of the
velocity $v$ (a constant of motion) and the amplitude of the  field $K$.
To illustrate the effect of varying them we have studied
the Poincar\'e section.
In Fig. (\ref{fig1}) we plot the surface section obtained
by using the plane $y=0$ for $K=1$ and several initial conditions.
We see that as $v$ tends to zero the trajectories are regular.
As $v$ increases, a smooth transition to a chaotic behavior is
observed. If one considers a chain of more than two sites of coupled maps of this
kind, one can expect  the property of chaoticity
to be  stronger for  larger system sizes.

\subsection{Numerical Analysis of the Fourier Law}

To compute the heat conductivity one has two possible
procedures: the first is the direct non-equilibrium
measure with two reservoirs at different temperatures,
taking the ratio between the time averaged flux
and the temperature gradient.
If a linear response regime is applicable, one can also use
the Green-Kubo formula, which
enables to calculate transport coefficients as
integrals of autocorrelation function in equilibrium
states. The thermal
conductivity is then given by
\be
k(T) = \frac{N}{T^2}\sum_{t=0}^{\infty}
<J(t)J(0)>_{GB} (1-\frac{\delta_{t,0}}{2})
\ee
where $J(t)$ is the flux per particle
\be
J(t) = \frac{1}{N}\sum_{i=1}^N J_i(t)
\ee
The local heat flux $J_i(t)$ is  given by
the change in energy for the couple $(i,i+1)$
before and after a kick.
The angular brackets denote here an equilibrium
average and the factor with the Kronecker
delta function arise from the  discreteness
of time.

\begin{figure}
\begin{center}
\includegraphics[angle=-90,width=7.cm]{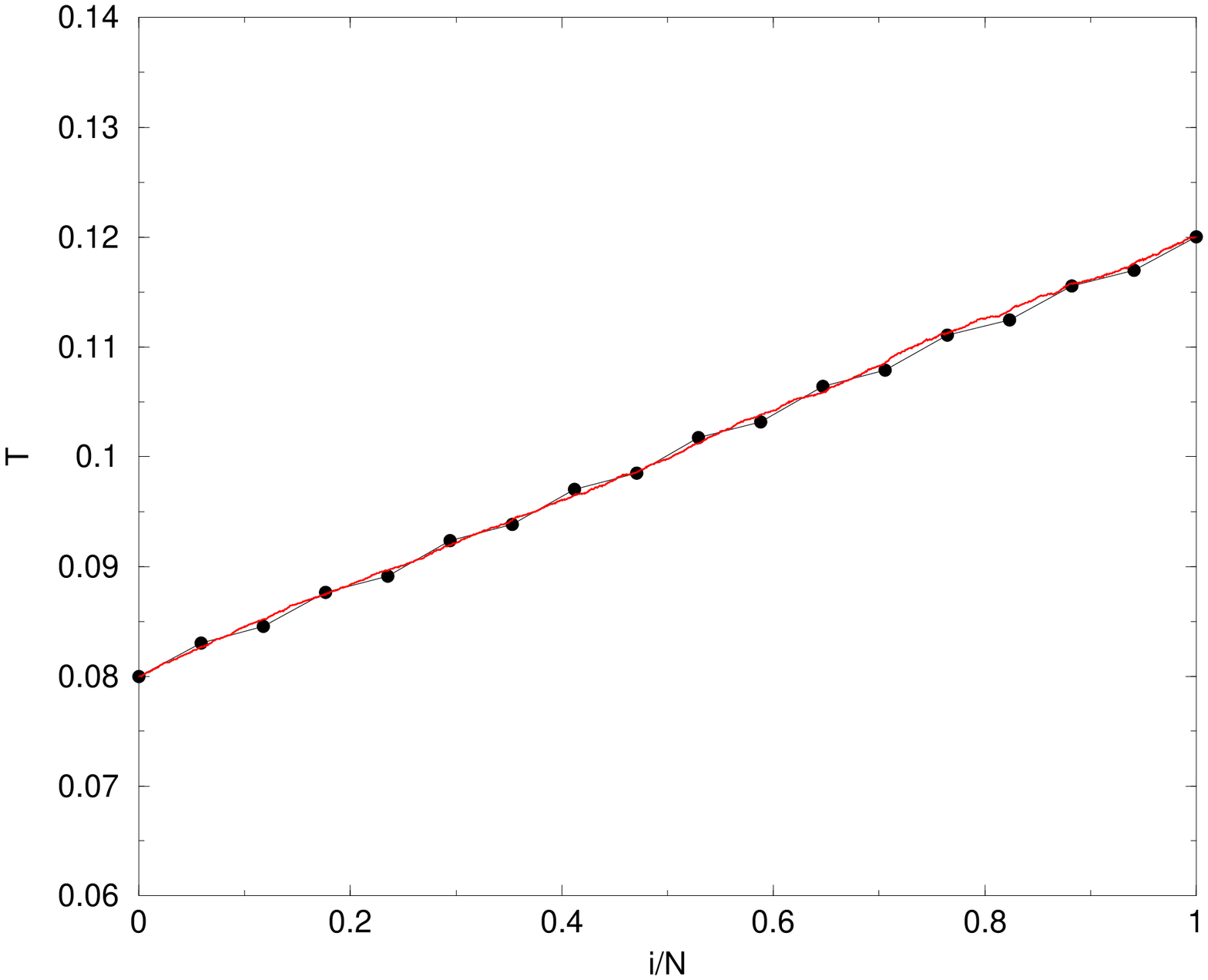}
\hspace{1.cm}
\includegraphics[angle=-90,width=7.cm]{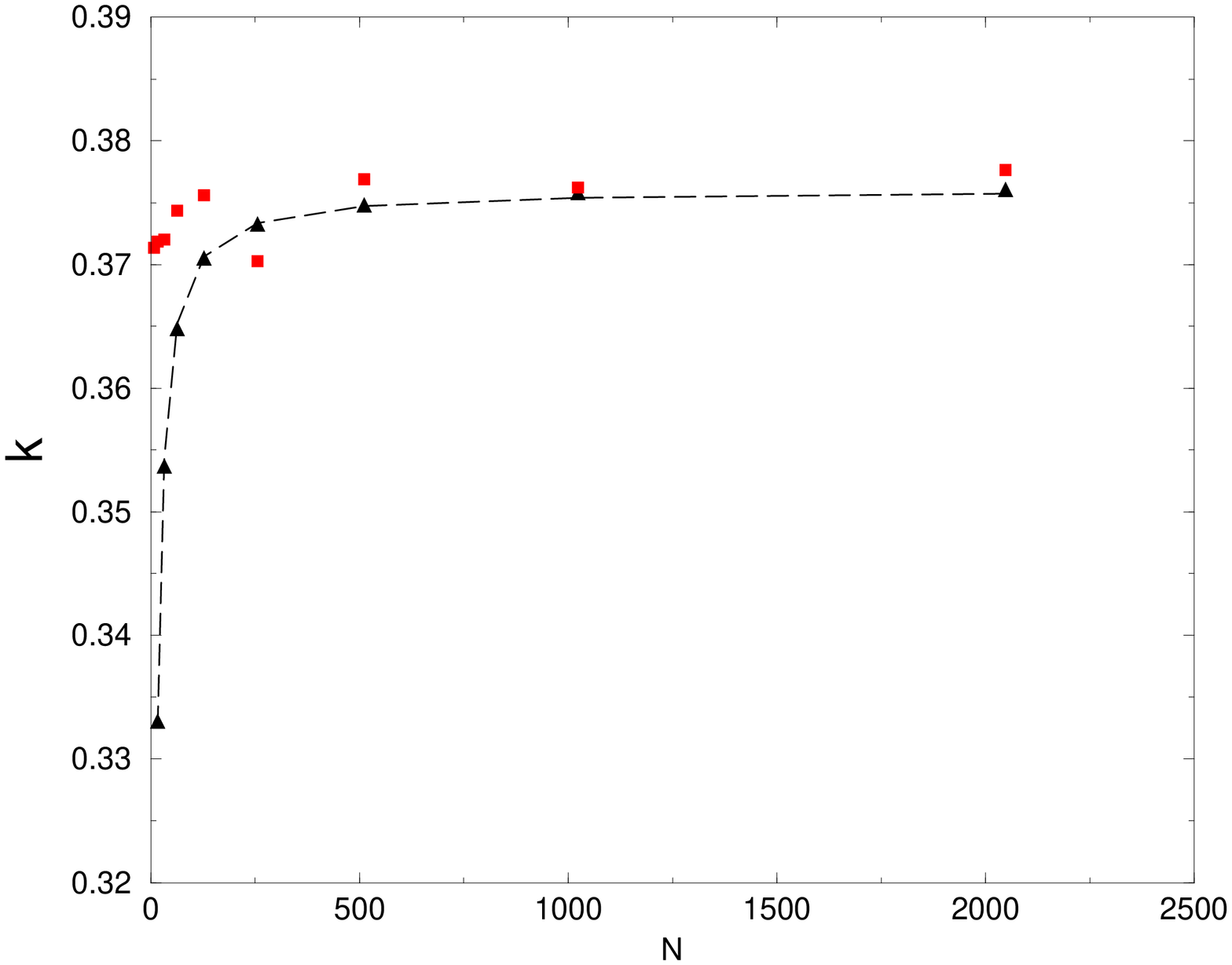}
\caption{{\small
Results of simulations for $K=1$.
Left: temperature profile with two heat baths for $N=16$ (circles)
and $N=2048$ (full line).
Right: Heat conductivity versus the chain length $N$
obtained throught non equilibrium simulations (squares)
and GK formula (triangles). The continous line represent
the best fit with a function $k_{\infty}+ a/N$.}}
\end{center}
\label{fig2}
\end{figure}

We first performed the numerical simulation with
two heat baths at temperatures $T_L = 0.08$ and $T_R = 0.12$.
After a transient of $10^7$ steps, we checked that a linear
temperature profile is established by measuring the temporal
average of twice the kinetic energy at each site
(see Fig. (\ref{fig2})-left).
We also kept track of the time averaged
flux until it converged to its stationary value. We used at least $10^8$
steps to check that fluctuations are less then a few percent.
The measured heat conductivity is reported in Fig. (\ref{fig2})-right.
for different chain lengths from $N=8$ to $N=2048$.
One can see that the conductivity has some fluctuation
in the interval $[0.37,0.38]$ with an overall trend to stabilize,
which suggests the approach to a finite value in the
thermodynamic limit.
We ran the dynamics for a large number of different initial
random conditions (at least $10^3$) and computed the heat
conductivity by the Green-Kubo formula.
We chose $e=E/N=0.05$ which, according to the virial
theorem, gives a kinetic temperature $T=0.1$, the average
value of the temperature of the two baths.
The best fit of the data shows a convincing evidence
that finite size effect are of the form ${\cal O}(1/N)$
with an asymptotic value $k_{\infty}= 0.376$.

\subsection{High temperature limit of the discrete model}

In the high temperature limit the velocities $\dot x_i$ are large so that between two
consecutive kicks new positions are translated
by a large amount. Since the spatial coordinates are taken modulo $2\pi$,
the sequence of  positions constitutes a (quasi) random number generator.
Hence, we assume that in this limit, the position of the particle when the fields
are flashed can be taken as  uniformly randomly distributed.
Because the magnetic fields are functions of the positions, this  means
that the fields themselves are random, with a distribution that is easily derived
from the position dependence of the fields.
All in all, what we have is that the velocity vector turns at each kick by a random amount,
the probability distribution of which is known.
Let us consider  the change in energy for the couple $(i,i+1)$
before ($e_i = \frac{1}{2} v_i^2$) and after ($e_i' = \frac{1}{2} v_i^{'2})$ a kick.
\bea
e'_{i}   & = &  c^2(B_i) \,e_{i}  + s^2(B_i)\, e_{i+1}
                + 2 \, s(B_i) \,c(B_i) \,\sqrt{e_i e_{i+1}}
\nonumber \\
e'_{i+1} & = &  s^2(B_i) \,e_{i}  + c^2(B_i)\, e_{i+1}
                - 2 \, s(B_i) \,c(B_i) \,\sqrt{e_i e_{i+1}}
\eea
where, of course, $e'_{i} + e'_{i+1} = e_{i} + e_{i+1}$.
Suppose now that $x_i,x_{i+1}$ are well approximated by
independent uniform random process on the interval $[0,2\pi]$.
Then the magnetic field (\ref{magneticfield}) will be
a random variable and, with this particular choice for
$G$, its probability distribution in the interval $[-2 K\pi,2 K\pi]$
will be
\be
p(B) = \left\{
\begin{array}{ll}
\frac{1}{2K\pi}\left(1 + \frac{1}{2K\pi} x\right) & x \in [-2K\pi,0] \\
\frac{1}{2K\pi}\left(1 - \frac{1}{2K\pi} x\right) & x \in [0,2K\pi] \\
\end{array}
\right.
\ee
The mean energy of a couple of particles will be
redistributed according to the rule:
\bea
\label{rule}
<e'_{i}>   & = &  x_K   \,<e_{i}> + \, (1-x_K)\;<e_{i+1}> \nonumber \\
<e'_{i+1}> & = & (1-x_K)\;<e_{i}> + \, x_K    \,<e_{i+1}>
\eea
since
\bea
<c^2 (B_i)>  & = &
\int_{-2K\pi}^{2K\pi} \cos^2(B)p(B)dB \; = \;
\frac{1}{2} \left (1 + \left(\frac{\sin(2K\pi)}{2K\pi}\right)^2 \right) \; = \;
x_{K}
\nonumber \\
<s^2 (B_i)>  & = &
\frac{1}{2} \left (1 - \left(\frac{\sin(2K\pi)}{2K\pi}\right)^2 \right) \; = \;
 1-x_{K}
\nonumber \\
<s(B_i) c(B_i)>  & =  &  0
\eea
Here expectation values are taken over an ensemble of systems having the same velocities but
random positions. If we choose $K$ to be an integer or a half-integer,
the dynamical rule (\ref{rule}) becomes
({\em only as far as the means are concerned})
the one of the stochastic model introduced by
Kipnis, Marchioro and Presutti
\cite{GKMP:81,KMP:82,EY:04}, where the total energy
of a particle pair is equally redistributed
between the two $(x=1/2)$.

Let us  give here a short computation of the heat
conductivity in the case $K=1$.
The idea is to use first self-averaging with respect
to the noise and then stationarity.
The local `temperature' is defined as twice the
kinetic energy $T_i(t) = \; <v_i^2(t)>$.
Using (\ref{one})-(\ref{two}) and imposing stationarity
$T_i(t) = T_i(t+1) = T_i$
one obtains the following recursion relation:
\bea
T_1 & = & T_L \nonumber \\
-T_{i-2} + 2T_i -T_{i+2} & = & 0 \quad\quad\;\; i=2,4,6,\ldots,N-2
\nonumber \\
T_{i+1} & = & T_{i}  \quad\quad i=0,2,4,6,\dots,N
\nonumber \\
T_N & = & T_R
\label{system}
\eea
This can be easily checked starting from a configuration
$(T_L,T_2,T_2,T_4,T_4,...,T_{N-2},T_{N-2},T_R)$,
making it evolve through the two kicks,
and demanding that the same configuration is recovered.
From the solution of the system (\ref{system}),
a linear temperature profile is obtained (apart from the
fact that the sites are in pairs):
\be
\label{temp-profile}
T_i = T_L + \frac{(T_R - T_L)}{N} i
\ee
It is easy to write a time-dependent version of (\ref{system}),
and check that the convergence to this profile is rapid, independently
of the length of the chain, as it is essentially a diffusion equation.

\noindent
Between $t$ and $t+ 1/2$ we consider the local flux
of the odd sites, while between $t+ 1/2$ and
$t + 1$ we consider the local flux on the even sites.
Self-averaging plus stationary gives
\bea
j_0 & = & \frac{1}{2} (T_L + T_1) - T_L \nonumber \\
j_i & = & \frac{1}{2} (T_i + T_{i+1}) - T_i
\quad\quad i=1,2,\ldots,N-1 \nonumber \\
j_N & = & \frac{1}{2} (T_R + T_N) - T_N
\eea
The solution is an average  site-independent local flux
\be
j_i = \frac{1}{2} \frac{T_R - T_L}{N}
\ee
and a spatial average flux
\be
\label{flux}
J = \frac{1}{N} \sum_{i=0}^{N}j_i = \frac{1}{2} \frac{T_R - T_L}{N}
\ee

\noindent
Putting together (\ref{temp-profile}) and (\ref{flux})
we verify Fourier law $J = k \nabla T$ with an heat
conductivity $k= 1/2$.

\begin{figure}
\label{fig3}
\begin{center}
\includegraphics[width=9.cm,angle=-90]{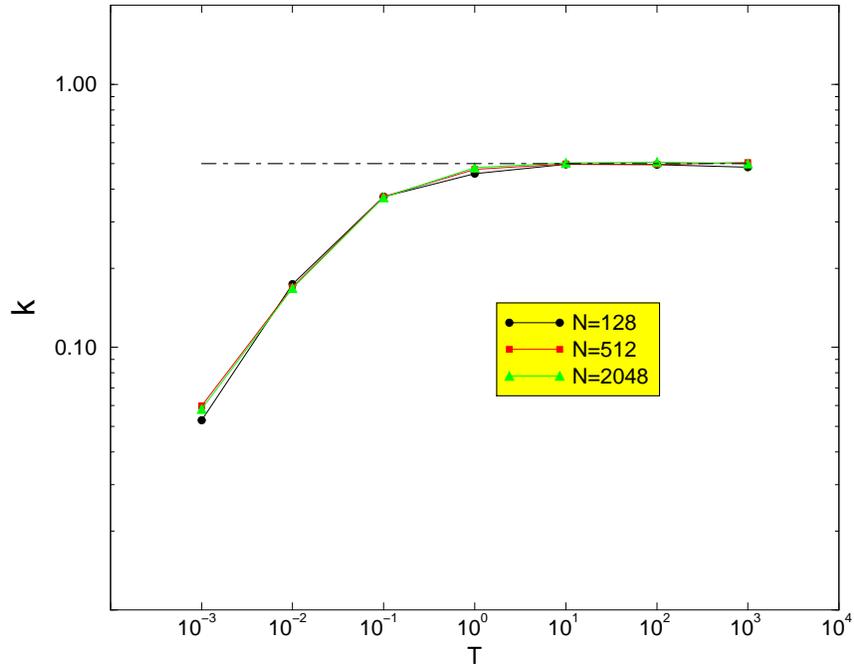}
\caption{{\small Thermal conductivity $k$ versus the temperature
(note the log-log scale). The dashed line corresponds to the value
$k=1/2$. Results are plotted for three different
$N$ values (see legend). Here only a single trajectory
has been used in the application of the Green-Kubo formula.
However the good overlap between the three curves indicates
reliability of the computation.}}
\end{center}
\end{figure}

\noindent
In Fig. (\ref{fig3}) we report the result of microcanonical
simulations at different temperatures. One can see that for
temperatures $T\geq 10$ the heat conductivity is basically
constant and its numerical value coincides with the value
we have just calculated in the high temperature regime.

\subsection{Momentum-conserving model}

We now run the arguments for a momentum-conserving model.
Consider a coupling between three sites which is given by a
magnetic field in the direction (1,1,1).
The analogue of (\ref{cosa}) for this case is:
\be
R_i(t) = \frac{1}{3}\left (
\ba{ccc}
1 + 2 \, c(B_i) &
1 - c(B_i) + \sqrt{3}\, s(B_i) &
1 - c(B_i) - \sqrt{3}\, s(B_i)
\\
1 - c(B_i) - \sqrt{3}\, s(B_i) &
1 + 2 \, c(B_i) &
1 - c(B_i) + \sqrt{3}\, s(B_i)
\\
1 - c(B_i) + \sqrt{3}\, s(B_i) &
1 - c(B_i) - \sqrt{3}\, s(B_i) &
1 + 2 \, c(B_i)
\ea
\right )
\ee
where now
\bea
c(B_i) & = & \cos (\sqrt{3} B_i(t))
\nonumber \\
s(B_i) & = &\sin (\sqrt{3} B_i(t))
\eea
A simple choice for $B_i = G(x_{i},x_{i+1},x_{i+2})$ is:
\begin{equation}
G(x_{i},x_{i+1},x_{i+2}) = K[(x_{i} + x_{i+1} + x_{i+2}) -3\pi]
\label{field3}
\end{equation}
Clearly, the velocity in the (1,1,1) direction (the sum of the three velocities)
is not affected.
If we now alternate kicks which couple sites $(1,2,3)$, $(4,5,6)$, $(7,8,9)$ ,..., etc,
with kicks which couple  sites $(2,3,4)$, $(5,6,7)$, $(8,9,10)$, ..., we obtain a map
that at each kick conserves the total sum of the velocities
(and, of course, the energy).

\noindent
We can repeat the argument in the previous section to obtain the high energy limit.
We set the multiplicative constant $K$ to $1/\sqrt{3}$, for simplicity.
If we consider that at each step the positions $x_i,x_{i+1},x_{i+2}$ are uniform and independent
random with the choice (\ref{field3}) we have that:
\begin{equation}
\langle c(B_i) \rangle = \langle s(B_i) \rangle = \langle c(B_i) s(B_i) \rangle = 0
\end{equation}
\begin{equation}
\langle c^2(B_i) \rangle = \langle s^2(B_i) \rangle = \frac{1}{2}
\end{equation}
and we see that the average energies evolve during a kick as:
\be
\left (
\ba{c}
\langle e_i' \rangle
\\
\langle e_{i+1}' \rangle
\\
\langle e_{i+2}' \rangle
\ea
\right )
 = \frac{1}{3}\left (
\ba{ccc}
1 &
1 &
1
\\
1 &
1 &
1
\\
1 &
1 &
1
\ea
\right )
\left (
\ba{c}
\langle e_i \rangle
\\
\langle e_{i+1} \rangle
\\
\langle e_{i+2} \rangle
\ea
\right )
\ee


\noindent
Again we can easily find an equation for the $T_i$, and check that the profile is linear,
with the only difference that now the sites come in triplets:
\be
(T_L,T_L,T_3,T_3,T_3,T_6,T_6,T_6,\ldots,T_{N-3},T_{N-3},T_{N-3},T_R)
\ee
and the temperatures satisfy
\be
-T_{i-3} + 2T_i -T_{i+3} =  0 \quad\quad\;\; i=3,6,9,\ldots,N-3
\ee
From this we easily obtain a value for the thermal conductivity
$k=1$.

\begin{figure}
\label{fig4}
\begin{center}
\includegraphics[width=9.cm,angle=-90]{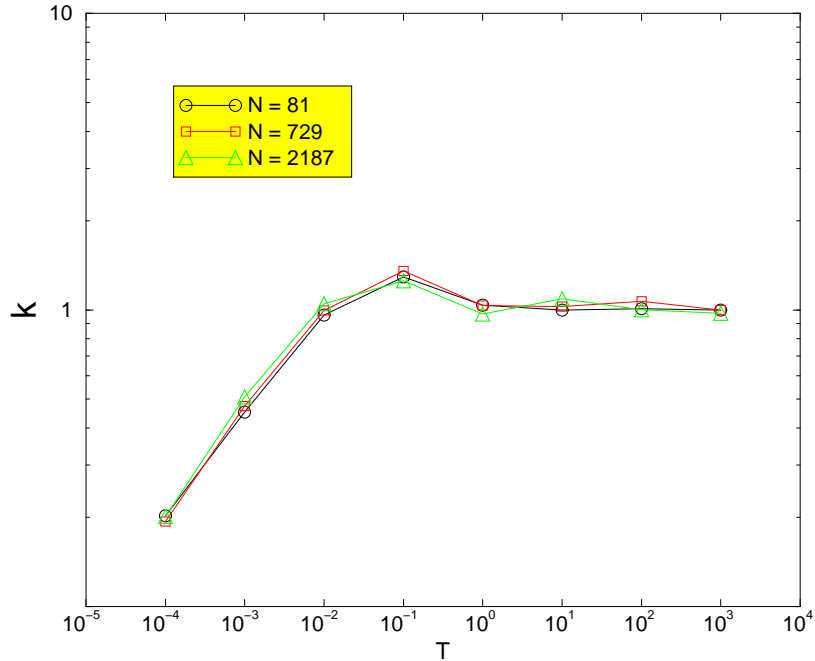}
\caption{{\small
Same as in Fig.(\ref{fig3}) but for the momentum conserving case. The analytical high-temperature
 value is $k=1$. }}
\end{center}
\end{figure}

The numerical simulations do agree with this value in the
high temperature regime and give a finite heat
conductivity at each temperature, see Fig. \ref{fig4}

\section{A stochastic minimal model}

As was done in the previous section for
the discrete time model we study here a
high temperature limit which becomes  a continuous time model,
by replacing the magnetic fields with suitable  random
processes.

\subsection{Fokker-Planck equation}

In the limit in which the field is very weak, and the velocities (and hence the energy)
are very large,
we get a velocity field that is randomly exchanging momentum, but in small amounts
each step.
The deterministic equations of motion
(\ref{newton}) can then be replaced by
a system of Langevin equations with
multiplicative noise
\be
\label{langevin}
\frac{dv_i}{dt}  =  \sum_{j=1}^N B_{ij} v_j
\ee
with the $B_{ij}$ of the nearest-neighbour form (\ref{b}):
\be
B_{ij}= B_i(t) (\delta_{i,j+1}-\delta_{i+1,j})
\ee
and $B_i$ a white, Gaussian random variables with
\footnote{This being a multiplicative process, the convention (Ito, Stratonovitch...)
should be specified. In fact, the precise definition will be implicit in the
Fokker-Planck equation (\ref{uno})(\ref{dos})}
\be
<B_j(t)> \;= 0 \quad\quad\quad
<B_j(t) B_k(t')> = 2 \;\delta_{jk}\;\delta(t-t')
\ee
Standard techniques \cite{Ri} yield the following
Fokker-Planck equation for the evolution
on the velocities probability distribution
\be
\frac{\p P}{\p t} = - \sum_{i=1}^N L_{i,i+1}^2 P\;
\label{uno}
\ee
where the operators $L_{i,i+1}$ are
`angular momentum' operators corresponding to the Laplacian on the sphere:
\be
L_{i,i+1} = v_i \,\frac{\p}{\p v_{i+1}} -
v_{i+1} \,\frac{\p}{\p v_{i}}
\label{dos}
\ee
If we add to the bulk system the interaction with
two heat baths connected to the first and last
particle (respectively at temperatures $T_L$ and $T_R$)
and represented as Ornstein-Uhlenbeck processes,
we arrive at
\be
\label{fp2}
\frac{\p P}{\p t} = \sum_{i=1}^N L_{i,i+1}^2 P\;
+ \p_1(v_1 P) + T_L \p_1^2 P \;
+ \p_N(v_N P) + T_R \p_N^2 P
\ee
This diffusion process has also been considered by
Olla \cite{Olla}, in combination with  a deterministic harmonic part.

One can also easily write a momentum-conserving variant. Indeed, considering the
low-field version of the conserving map, we get:
\be
\frac{\p P}{\p t} = - \sum_{i=1}^N \{ L_{i,i+1} + L_{i+1,i+2} + L_{i+2,i}\}^2 P\;
\label{tres}
\ee
where in a closed chain the indices are understood modulo $N$.

\subsection{Expectation values}

We would like to compute the energy and flux expectation values
and also energy correlations. One may derive an equation of
motion for these quantities directly from the Fokker-Planck
equation. Multiplying for instance (\ref{fp2}) by $v_i^2$
and integrating the resulting equations, i.e.
\bea
\frac{\p}{\p t} \int v_i^2 P(v) d^Nv & = &
\int v_i^2 \left (\sum_{i=1}^N L_{i,i+1}^2 P(v) \right) d^Nv \nonumber \\
& + &
\int v_i^2 \left( \p_1(v_1 P(v)) + T_L \p_1^2 P(v)\right) d^Nv \nonumber \\
& + &
\int v_i^2 \left(
\p_N(v_N P(v)) + T_R \p_N^2 P(v) \right ) d^Nv
\eea
by using repeatedly integration by parts
(with $v_i\p_i = \p_i v_i -1$)
we obtain
\bea
\frac{d}{dt} <v_1^2> & = &
-4 <v_1^2>  + 2 <v_2^2> + 2 T_L \nonumber \\
\frac{d}{dt} <v_i^2> & = &
2 <v_{i-1}^2> -4 <v_i^2>  + 2 <v_{i+1}^2>
\quad\quad \mbox{for} \quad i = 2,\ldots,N-1 \nonumber \\
\frac{d}{dt} <v_N^2> & = &
-4 <v_N^2>  + 2 <v_{N-1}^2> + 2 T_R
\eea
In the stationary state $\frac{d}{dt} <v_i^2> = 0$
so that the temperature profile is obtained as the
solution on a linear system of equations.
The solution is:
\be
\label{temper-prof}
T_i = T_L + \frac{(T_R - T_L)}{N+1} \;i
\ee
In the stationary state there is a net heat
current $J$ flowing through the lattice.
This can be calculated by directly measuring
the energy exchange with the two baths.
The energy flux from the left reservoir to the
first particle is
\be
<J_1> \; = \; <v_1^2> - \; T_L
\ee
while the energy flux from the last particle to
the right reservoir is
\be
<J_N> \; =  \; <v_N^2> - \; T_R
\ee
In both cases, by using (\ref{temper-prof}) we obtain
\be
\label{jflux}
<J_1> \; = \; <J_N> \; = \; J = \frac{(T_R - T_L)}{N+1}
\ee
Putting together (\ref{temper-prof}) and (\ref{jflux})
we find that Fourier law holds for the stochastic
model with a heat conductivity $k=1$.

\subsection{The stationary measure}

We show here that the stationary measure in the presence of heat flow
cannot be of either Gaussian or product form.
First of all, the evolution operator is invariant with respect to change in sign of
any single velocity, and hence $P(v_1\ldots v_N)= P(T_1,...,T_N)$, i.e. the velocities
only appear as even powers.

\noindent
Proposing a Gaussian measure of the form:
\be
P(v_1\ldots v_N) \sim e^{-\frac{1}{2} \sum_{ij} A_{ij} v_i v_j}
\ee
substituting into (\ref{fp2}), we get:
\be
\left[ \sum_{ij} A_{ij} v_i v_j \right]^2  = \left[ \sum_{ij} A^2_{ij} v_i v_j\right] (\sum v_i^2)
\quad \forall v\in \RR^N
\ee
Going to the diagonal basis $A = A_i \delta_{ij}$, this equation implies $(A_i-A_l)^2=0$,
i.e., $A$ is proportional to the identity. This is impossible if there is heat flow.

\noindent
Let us see that a product measure:
\be
P(v_1\ldots v_N) = \prod_{i=1}^{N} p_i(v_i)
\ee
 is in general not possible if there is heat flow. In the stationary state we have
\be
\frac{L_{FP} P}{P} = \sum_{i=1}^N
\frac{L_{i,i+1}^2 p_i(v_i)p_{i+1}(v_{i+1})}{p_i(v_i)p_{i+1}(v_{i+1})}
=0
\ee
Because of their different arguments, each term must vanish separately, thus:
\be
(L_{i-1,i}^2 + L_{i,i+1}^2)p_{i-1}(v_{i-1})p_i(v_i)p_{i+1}(v_{i+1}) = 0
\ee
This is an equation of the form $(L_x^2+L_y^2)\psi=0$ (with $L_x$, $L_y$ the $SU(2)$ operators),
which can only be satisfied by the quantum numbers $(l,m)=(00)$, the spherically symmetric
function. In our case this means that:
\be
p_{1}(x)p_2(y)p_{3}(z) = g(x^2 + y^2 + z^2)
\ee
which is a product only if we have an isotropic Gaussian, again impossible in the
presence of heat flow.

The equation of motion for
$n$-point correlation functions of these models are linear and close within themselves.
This is a strong suggestion that the models may be integrable, but we shall not persue
this line here.

\section{Conclusions}

We have studied a family of models of heat transport whose high energy limit
is analytically solvable. The intermediate energies are easy to simulate,
the approach to the asymptotic value can be easily tested. The momentum-conserving
version of the model has finite conductivity, providing further confirmation
of the fact that in itself momentum conservation does not imply anomalous conductivity.

It would be nice in the future to investigate in the simple framework
of ``magnetic kicks'' we have introduced a system which has also
a potential energy. For example one could start
from a system of coupled linear oscillators, establishing a connection with the work of Olla \cite{Olla}.

Another interesting problem is to solve the Fokker-Planck equation
to find the stationary distribution of our system. This could shead some light
also on the stochastic model introduced by Presutti et al. \cite{KMP:82}.
There again the stationary measure is a product measure only locally
(i.e. in the approximation to first order in $L^{-1}$).

\vspace{0.5cm}
\noindent
{\bf Acknowledgements:} We wish to thank  P. Contucci, M. Degli Esposti, J-P. Eckmann, J. Lebowitz,
M. Lenci, S. Lepri, R. Livi, C. Mejía-Monasterio, S. Olla, A. Politi, E. Presutti and H.H. Rugh for some
helpful discussion and correspondence and L. Bertini for the preprint \cite{BGL:05}.
C.G.  acknowledges the continuous encouragement from S. Graffi and G. Jona-Lasinio.


\vspace{1.cm}
\begin{thebibliography}{X}



\bibitem{BLR:00}
{\sc F. Bonetto, J. Lebowitz, L. Rey-Bellet}, \, ``Fourier's Law:
a Challenge for Theorists'', \, in {\em Mathematical Physics 2000},
A. Fokas, A. Grigoryan, T. Kibble and B. Zegarlinsky (Eds.),
Imperial College, London, (2000), pp. 128-150.

\bibitem{LLP:03}
{\sc S. Lepri, R. Livi, A. Politi }, \, ``Thermal conduction in
classical low dimensional lattices'' , \, {\sl Phys. Rep.}
{\bf 377} (2003), 1

\bibitem{RLL:67}
{\sc Z. Rieder, J.L. Lebowitz, E. Lieb}, \, ``Properties
of a harmonic crystal in a stationary nonequilibrium state'', \,
{\sl J. Math. Phys.} {\bf 8} (1967), 1073

\bibitem{N:70} {\sc H. Nakazawa},\, ``On the lattice thermal
conduction'', \, {\sl Suppl. Progr. Theor. Phys.} {\bf 45}
(1970), 231-262

\bibitem{LLP:97}
{\sc S. Lepri, R. Livi and A, Politi}, \, ``Heat conduction in
chains of nonlinear oscillators'' , \, {\sl Phys. Rev. Lett. } {\bf
78} (1997), 1896

\bibitem{H:99}
{\sc T. Hatano}, \,
``Heat conduction in the diatomic Toda lattice revisited'', \,
{\em Phys. Rev. E} {\bf 59},(1999) R1

\bibitem{D:01}
{\sc A. Dhar} \,
``Heat Conduction in a One-Dimensional Gas of Elastically
Colliding Particles of Unequal Masses'', \,
{\em Phys. Rev. Lett.} {\bf 86}  (2001) 3554

\bibitem{sym} The same situation arises in Refs. 
 \cite{MLL:01} \cite{EY:04}, where the rotation of
the velocity vector is brought about by the impact on a `rough' cylinder. 
 
\bibitem{GNY:02}
{\sc P. Grassberger, W. Nadler and L. Yang} \,
``Heat Conduction and Entropy Production in a
One-Dimensional Hard-Particle Gas'',\,
{\em Phys. Rev. Lett.} {\bf 89} (2002) 180601

\bibitem{CP:03}
{\sc G. Casati, T. Prosen} ,\,
``Anomalous Heat Conduction in a Di-atomic One-Dimensional Ideal Gas'',\,
{\em Phys. Rev. E} {\bf 67}, (2003) 015203(R)

\bibitem{DN:03}
{\sc J. M. Deutsch and O. Narayan} \,
``One-dimensional heat conductivity
exponent from a random collision model'', \,
{\em Phys. Rev. E}  {\bf 68}, (2003) 010201

\bibitem{GLPV:00}
{\sc C. Giardin\'a, R. Livi, A. Politi and M. Vassalli}, \, ``Finite
thermal conductivity in 1D lattices'' , \, {\sl Phys. Rev. Lett.}
{\bf 84 } (2000), 2144

\bibitem{GS:00}
{\sc O. V. Gendelman and  A. V. Savin}, \,
``Normal Heat Conductivity of the One-Dimensional Lattice
with Periodic Potential of Nearest-Neighbor Interaction'', \,
{\sl Phys. Rev. Lett.} {\bf 84}, (2000) 2381

\bibitem{CFVV:84}
{\sc G. Casati, J. Ford, F. Vivaldi and W. Visscher} \,
``One-Dimensional Classical Many-Body System
Having a Normal Thermal Conductivity'', \,
{\sl Phys. Rev. Lett.} {\bf 53} (1984), 1120

\bibitem{PR:92}
{\sc T. Prosen and M. Robnik} \,
``Energy transport and detailed verification of
Fourier heat law in a chain of colliding harmonic oscillators'', \,
{\sl J. Phys. A: Math. Gen.} {\bf 25} (1992) 3449-3472

\bibitem{HLZ:98}
{\sc B. Hu, B. Li and H. Zhao} \,
``Heat conduction in one-dimensional chains'', \,
{\em Phys. Rev. E} {\bf 57} (1998) 2992

\bibitem{TBSZ:99}
{\sc G. P. Tsironis, A. R. Bishop, A. V. Savin and A. V. Zolotaryuk} \,
``Dependence of thermal conductivity on discrete breathers
in lattices'',\,
{\em Phys. Rev. E} {\bf 60} (1999) 6610

\bibitem{BS:81}
{\sc L. Bunimovich and Ya. G. Sinai} \,
{\sl Comm. Math. Phys.} {\bf 78} (1981) 661

\bibitem{LS:78}
{\sc J. L. Lebowitz and H. Spohn} \,
{\sl J. Stat. Phys.} {\bf 19} (1978) 633

\bibitem{AACG:99}
{\sc D. Alonso, R. Artuso, G. Casati and I. Guarneri} \,
``Heat Conductivity and Dynamical Instability'',\,
{\em Phys. Rev. Lett.} {\bf 82} (1999) 1859


\bibitem{MLL:01}
{\sc C. Mejía-Monasterio, H. Larralde and F. Leyvraz} \,
``Coupled Normal Heat and Matter Transport in a Simple Model System'',\,
{\em Phys. Rev. Lett.} {\bf 86} (2001) 5417

\bibitem{ARV:02}
{\sc D. Alonso, A. Ruiz and I. de Vega} \,
``Polygonal billiards and transport: Diffusion and heat conduction'',\,
{\em Phys. Rev. E} {\bf 66} (2002) 66131

\bibitem{LWH:02}
{\sc B. Li, L. Wang and B. Hu} \,
``Finite Thermal Conductivity in 1D Models Having Zero
Lyapunov Exponents'',\,
{\em Phys. Rev. Lett.} {\bf 88} (2002) 223901

\bibitem{LCW:03}
{\sc B. Li, G. Casati and J. Wang} \,
``Heat conductivity in linear mixing systems'',\,
{\em Phys. Rev. E} {\bf 67} (2003) 21204

\bibitem{CL:71}
{\sc A. Casher, J.L. Lebowitz} \,
{\em J. Math. Phys.} {\bf 12} (1971) 1701

\bibitem{RG:71}
{\sc R. J. Rubin, W. L. Greer} \,
{\em J. Math. Phys.} {\bf 12} (1971) 1686

\bibitem{OL:74}
{\sc A. J. O' Connor, J. L. Lebowitz} \,
{\em J. Math. Phys.} {\bf 15} (1974) 692

\bibitem{LZH:01}
{\sc B. Li,H. Zhao, B. Hu} \,
``Can Disorder Induce a Finite Thermal Conductivity in 1D Lattices?''\,
{\em Phys. Rev. Lett.} {\bf 86} (2001) 63

\bibitem{GKMP:81}
{\sc A. Galves, C. Kipnis, C. Marchioro and E. Presutti} \,
{\em Comm. Math. Phys.} {\bf 81} (1981) 127

\bibitem{KMP:82}
{\sc C. Kipnis, C. Marchioro and E. Presutti} \,
``Heat flow in an exactly solvable model'', \,
{\em J. Stat. Phys.} {\bf 27} (1982) 65

\bibitem{EY:04}
{\sc J. P. Eckmann and L-S. Young } \,
``Temperature profiles in Hamiltonian heat conduction'', \,
{\em Europhys. Lett.} {\bf 68} (2004) 790-796, see also
{\sc  C. Mejia-Monasterio, H. Larralde, F. Leyvraz}\,
``Observation of coupled normal heat and matter transport in a simple model system'',
{\em  Phys. Rev. Lett.} 86 (2001) 5417-5420

\bibitem{BGL:05}
{\sc L. Bertini, D. Gabrielli, J.L. Lebowitz} \,
``Large deviations for a stochastic model of heat flow'', \,
preprint (2005), to be published.

\bibitem{LLP:98}
{\sc S. Lepri, R. Livi and A, Politi}, \, ``On the anomalous thermal
conductivity in one dimensional lattices'' , \, {\sl Europhys. Lett.
} {\bf 43 } (1998), 271

\bibitem{NR:02}
{\sc O. Narayan and S. Ramaswamy}, \, ``Anomalous Heat Conduction
in One-Dimensional Momentum-Conserving Systems'', \,
{\sl Phys. Rev. Lett.} {\bf 89} (2002), 200601

\bibitem{cinesi}
{\sc J-S. Wang and B. Li}\,
``Mode-coupling theory and molecular dynamics simulation for
heat conduction in a chain with transverse motions'',\,
{\em Phys. Rev. E} {\bf 70} (2004) 021204

\bibitem{LL:00}
{\sc A. Lippi, R. Livi }, \,
``Heat conduction in 2d nonlinear lattices'' , \,
{\sl J. Stat. Phys. } {\bf 100 } (2000), 1147

\bibitem{GY:02}
{\sc P. Grassberger and L. Yang}, \,
``Heat Conduction in Low Dimensions: From Fermi-Pasta-Ulam Chains
to Single-Walled Nanotubes''; Preprint {\em cond-mat/0204247} (2002)

\bibitem{Y:02}
{\sc Lei Yang} \,
``Finite Heat Conduction in a 2D Disorder Lattice'', \,
{\em Phys. Rev. Lett.} {\bf 88} (2002) 094301

\bibitem{Ri}
{\sc H.Risken}
{\em The Fokker Planck Equation}, \,
Springer-Verlag Berlin Heidelberg , 2nd Edition,
(1996)

\bibitem{Olla} S. Olla, to be published.


\end{thebibliography}
\end{document}